\documentclass[useAMS,usenatbib]{mn2e}

\usepackage{times} 
\usepackage{amsmath}
\usepackage{relsize}
\usepackage{epstopdf}
\usepackage{graphicx}
\usepackage{graphics}
\usepackage{multirow}
\usepackage{booktabs}
\usepackage{lscape}
\usepackage{longtable}
\usepackage[figuresright]{rotating}
\usepackage{multicol}
\usepackage{natbib}
\usepackage{rotating}
\usepackage{epsfig}
\usepackage{amsfonts}
\usepackage{threeparttable}
\usepackage[english]{babel}
\usepackage{textcomp}
\graphicspath{{/pc71disk/asb/INTData/LaTeX}}

\usepackage{natbib}
\usepackage{subfigure}
\usepackage{tabularx}
\title[The LDB age of the BPMG]{A lithium depletion boundary age of 21 Myr for the Beta Pictoris moving group}
\author[A. S. Binks \& R. D. Jeffries]{A. S. Binks\thanks{E$-$mail: a.s.binks@keele.ac.uk} and R. D. Jeffries\\
Astrophysics Group, School of Chemistry and Physics, Keele University, Keele, Staffordshire ST5 5BG}
\begin{document}

\date{Submitted 2 September 2013}

\pagerange{\pageref{firstpage}$-$$-$\pageref{lastpage}} \pubyear{2013}

\maketitle

\label{firstpage}

\begin{abstract}
Optical
spectroscopy is used to confirm membership
for 8 low-mass candidates in the young Beta Pic moving group
(BPMG) via their radial velocities, chromospheric activity and
kinematic parallaxes.
We searched for the presence of the Li\,{\sc i} 6708\AA\ resonance
feature and combined the results with literature measurements of other
BPMG members to find the age-dependent
lithium depletion boundary (LDB) -- the luminosity at which Li remains
unburned in a coeval group. The LDB age of the BPMG
is $21 \pm 4$\,Myr and insensitive to the choice of low-mass
evolutionary models. This age is more precise, likely to be more
accurate, and much older than that commonly assumed for
the BPMG. As a result, substellar and planetary
companions of BPMG members will be more massive than previously thought.

\end{abstract}


\section{Introduction}
The Beta Pic Moving Group (BPMG) is a young,
kinematically coherent group of a few dozen stars within $\sim 50$~pc of
the Sun 
\citep{1999Barrado-y-Navascues_et_al, 2001Zuckerman_et_al,
  2004Zuckerman_Song, 2008Torres_et_al}.  Members of moving groups
(MGs) like the BPMG represent some of the best observational targets
for advancing understanding of the early evolution of stars, their
circumstellar environments and planetary systems. They are
closer than equivalents in more spatially concentrated
clusters and star forming regions, offering better intrinsic
sensitivity and spatial resolution.  Young gas giant planets around MG
members are expected to be more luminous than in older systems,
and young stars are frequently surrounded by debris discs that may
evidence the formation of terrestrial planets or provide diagnostics of
unseen planets.  Studies that exploit the youth and proximity of BPMG
members include: the identification and direct imaging of a planet
(e.g. Lagrange et al. 2010, 2012; Bonnefoy et al. 2011, 2013) and
  debris disc (e.g. Smith \& Terrile 1984; Holland et al. 1998;
  Golimowski et al. 2006)
around its most luminous
member, the A0 star $\beta$ Pic; systematic surveys for planets, gas
and debris discs using high angular resolution imaging or infrared
diagnostics (e.g. Brandt et al. 2013; Wahhaj et al. 2013; 
  Dent et al. 2013); and testing low-mass stellar and substellar
evolutionary models using spatially resolved binary systems (Biller et al. 2010; Mugrauer et al. 2010).

\nocite{2010Mugrauer_et_al}
\nocite{2010Biller_et_al}
\nocite{2013Dent_et_al}
\nocite{2013Wahhaj_et_al}
\nocite{2013Brandt_et_al}
\nocite{2011Bonnefoy_et_al}
\nocite{2013Bonnefoy_et_al}
\nocite{2010Lagrange_et_al}
\nocite{2012Lagrange_et_al}
\nocite{1984Smith_Terrile}
\nocite{1998Holland_et_al}
\nocite{2006Golimowski_et_al}

All such investigations require age estimates for BPMG members,
either to place them in context with observations of other coeval
groups or to compare them with models for stars, discs and planets at
an absolute age -- for example to estimate masses (or
detection limits) of planetary companions from their
luminosities.  If BPMG membership is established,
then a common, coeval origin is usually assumed with other members, and the age
of the ensemble is estimated using the same techniques available for
young clusters.  The BPMG age is commonly assumed to
be $\sim$ 12\,Myr.  Estimates based on the positions of low-mass stars
in the Hertzsprung-Russell diagram (HRD) are $20\pm10$\,Myr
\citep{1999Barrado-y-Navascues_et_al} and $12^{+8}_{-4}$\,Myr
\citep{2001Zuckerman_et_al}.  Some estimates of kinematic ages, based
on the traceback of candidate members to their smallest volume, also
yield 10-12\,Myr with very small formal uncertainties
(Ortega et al. 2002, 2004; Song et al. 2003). However,
\citet{2007Makarov_et_al} finds a much less precise $22\pm 12$\,Myr.

\nocite{2002Ortega_et_al}
\nocite{2004Ortega_et_al}
\nocite{2003Song_et_al}

In this {\it Letter} we estimate an age for the BPMG based on the lithium
depletion boundary (LDB). In low-mass stars Li is burned rapidly
once PMS contraction raises core temperatures to $\sim
3\times10^{6}$~K. Convective mixing ensures that Li is then depleted
throughout the star, including the photosphere. The age at which
this occurs is mass-dependent and thus, in a coeval group, the
luminosity of a sharp transition between faint stars
that retain their initial Li and only slightly more luminous stars that
are entirely depleted, can be used to estimate the group age
\citep[e.g.][]{1997Bildsten_et_al}. In contrast with
ages determined from isochrone fitting in the HRD, the LDB technique
has little model dependence \citep{2004Burke_et_al} and has been 
used to estimate the ages of several young clusters
with ages $<40$\,Myr that are
comparable to the BPMG (NGC~1960 with age $22 \pm 4$\,Myr -
\cite{2013Jeffries_et_al}; IC~4665 with age $28\pm 5$\,Myr -
\cite{2008Manzi_et_al}; NGC~2547 with age $35\pm 3$\,Myr -
\cite{2005Jeffries_Oliveira}). New observations presented
here allow us to confirm membership for previously proposed low-mass BPMG candidates,
assess their Li content and determine an accurate LDB age for the
group.

\section{New low-mass candidates, observations, and data reduction}
\begin{figure*}
 \vspace{2pt}
\begin{center}
\includegraphics[width=\textwidth]{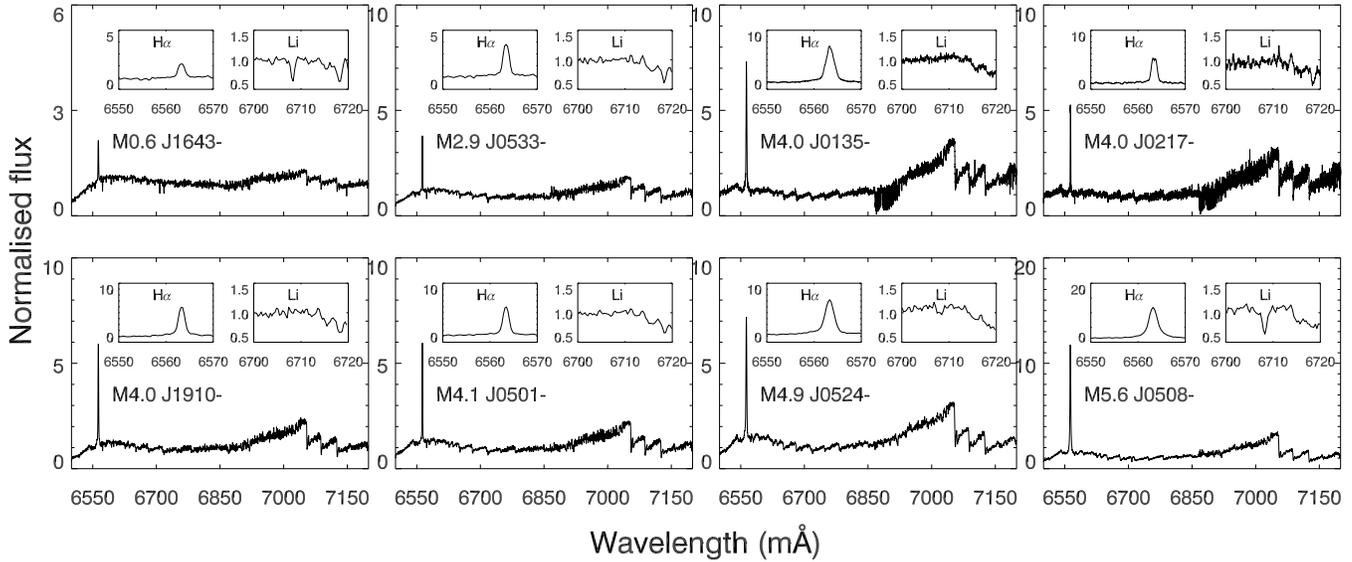}
\end{center}
 \caption{Spectra for the eight RV-confirmed members in our
   sample. Full 2MASS names are given in Table~1. The inserts 
   in each plot are normalised spectra in the
   regions of the H$\alpha$ and Li\,{\sc i} 6708\AA\ line.  All spectra
   (excluding objects `J0135-' and `J0217-', observed at the NOT, which
   have been blaze-corrected) have
   been subject to relative flux-calibration and telluric correction.
 }
  \label{fig:Spectra}
\end{figure*}

\begin{table*}
\centering
\caption{Confirmed BPMG members with their photometry (from UCAC4 --
  Zacharias et al. 2013;  2MASS -- Cutri et al. 2003), radial
  velocties (RV), difference between measured and predicted RVs (for group
  membership, $\Delta$RV), equivalent widths of H$\alpha$ and Li\,{\sc
    i}~6708\AA, distance (from a
  trigonometric parallax where available, otherwise the ``kinematic
  distance'' -- see Section~3.1) and a spectral type.}
\begin{tabular}{@{}llllllllllll@{}}
\toprule
Name & Ref & HJD & $V$ & $J$ & $K$ & RV & $\Delta$RV & H$\alpha$ EW & Li EW & distance & SpT \\
(2MASS) &  & 2450000+ & mag & mag & mag & kms$^{-1}$ & kms$^{-1}$ & \AA\ & m\AA\ & pc & M-  \\
\midrule

J01351393$-$0712517 & M13 &   6290.625    &     13.43  &      8.96  &      8.08  &       6.5  $\pm$       1.8$^{\textrm{d}}$  &       2.7  &      $-6.1$  &            $<$ 23 &      37.9  $\pm$       2.4$^{\textrm{k}}$  &       4.0  \\
J02175601$+$1225266 & S12 &    6291.572    &    14.09  &      9.96  &      9.08  &       7.0  $\pm$       1.3  &       1.4  &      $-4.2$  &           $<$ 22 &      67.9  $\pm$       6.1  &       4.0  \\
J05335981$-$0221325 & M13 &  6372.435    &    12.42  &      8.56  &      7.70  &      22.0  $\pm$       1.3  &       3.2  &      $-2.8$  &            $<$ 49 &      41.9  $\pm$       3.3  &       2.9  \\
J16430128$-$1754274 & M13 &  6375.736    &    12.57  &      9.44  &      8.55  &     $-10.0$  $\pm$       1.5$^{\textrm{e}}$  &       3.5  &      $-1.6$  &          364 $\pm$ $21^{\textrm{h}}$ &      59.2  $\pm$       2.8  &       0.6  \\
J05015881$+$0958587$^{\textrm{a}}$ & M13 &    6375.354    &    11.51  &      7.21  &      6.37  &      18.8  $\pm$       1.5$^{\textrm{f}}$  &       2.9  &      $-4.6$  &           $<$ 53$^{\textrm{i}}$ &      24.9  $\pm$       1.3$^{\textrm{l}}$  &       4.1  \\
J05241914$-$1601153$^{\textrm{b}}$ & M13 &  6376.383    &   13.57  &      8.67  &      7.81  &      20.6  $\pm$       4.1$^{\textrm{g}}$  &       0.0  &      $-7.0$  &          217 $\pm$ $29^{\textrm{j}}$ &      31.6  $\pm$       4.9  &       4.9  \\
J05082729$-$2101444 & M13 &  6377.361    &    14.41$^{\textrm{c}}$  &      9.72  &      8.83  &      22.8  $\pm$       3.8  &       2.0  &     $-12.9$  &          618 $\pm$ 43 &      30.5  $\pm$       4.9  &       5.6  \\
J19102820$-$2319486 & M13 &  6377.767    &    13.22  &      9.10  &      8.22  &      $-8.0$  $\pm$       1.7  &       3.8  &      $-4.8$  &            $<$ 55 &      69.2  $\pm$       3.4  &       4.0  \\

\bottomrule
\end{tabular}
\label{table:BPMG_Members1}
\begin{tablenotes}
\item [a]  \textbf{Notes.}
Targets from M13 = \cite{2013Malo_et_al}, S12 = \cite{2012Schlieder_et_al}.
Unresolved binary systems: (a) M3.0$+$M3.0 and (b)
M4.0$+$M4.5 \citep{2013Malo_et_al}.  (c) Photometry from the
NOMAD Catalog \citep{2005Zacharias_et_al}, $\sigma_{V} = 0.3$ assumed.
RVs previously measured: (d) 11.7 $\pm$ 5.3 $\textrm{kms}^{-1}$
\citep{2012Shkolnik_et_al}, (e) $-11.3$ $\pm$ 3.5 $\textrm{kms}^{-1}$
(Zwitter et al. 2008), (f) 14.9 $\pm$ 3.5 $\textrm{kms}^{-1}$
\citep{2007Kharchenko_et_al}, (g) 20.5 $\pm$ 4.0 $\textrm{kms}^{-1}$
\citep{2013Malo_et_al}.  Li EW values previously reported: (h)
300\,m\AA\ (Kiss et al. 2011), (i) 0 \citep{2009da-Silva_et_al}, (j)
223\,m\AA\ \citep{2013Malo_et_al}.  Distances from
trigonometric parallaxes: (k) \cite{2012Shkolnik_et_al} and (l)
\cite{2013Malo_et_al}; we derive kinematic distances of 37.7 $\pm$ 4.3 and 33.2 $\pm$ 3.7~pc, respectively.
\end{tablenotes}
\end{table*}

\nocite{2003Cutri_et_al}
\nocite{2008Zwitter_et_al}
\nocite{2011Kiss_et_al}
\nocite{2013Zacharias_et_al}

Indications of photospheric Li
abundance using the Li~{\sc
  i}~6708\AA\ line are available for a few low-mass BPMG members,
but are sparse in the critical region close to the expected LDB
location at 10-30\,Myr (spectral types M4--M5 --
Mentuch et al. 2008; Yee \& Jensen 2010).  We have
obtained new spectroscopy of M-type BPMG candidates with brightnesses
of  $10.6<V<14.6$, and spectral-type M ($V-K > 4$).
Targets were taken from probable members listed in the
proper-motion-based surveys of \citet{2012Schlieder_et_al},
\citet{2012Shkolnik_et_al}  and \citet{2013Malo_et_al}.  The spectroscopy is used to confirm kinematic
membership with a radial velocity measurement and ascertain the
presence or not of Li.  Spectroscopy was also obtained for a
sample of M dwarf candidates in the AB Doradus MG. The
full details of all targets and measurements will be presented in a
subsequent paper; in this {\it Letter}, we focus on BPMG candidates that
emerged as likely members. The results of the analyses for 
these stars are summarised in Table~1.

\nocite{2008Mentuch_et_al}
\nocite{2010Yee_Jensen}

Targets were observed on 28-29 December 2012 using
the 2.56-m Nordic Optical Telescope (NOT) and Fibre-fed Echelle
Spectrograph (FIES, R $\sim$ 46000), calibrated with simultaneous ThAr lamp
spectra.  Spectra covering the range $\lambda\lambda$3630-7260\AA\
were flat-fielded, extracted, wavelength calibrated and blaze-corrected 
using {\sc fiestool} \citep{stempels05}.  A second observing run
on 20-26 March 2013 used the 2.5-m Isaac Newton
Telescope (INT) and Intermediate Dispersion Spectrograph.  
The H1800V grating and a 1.4 arcsec
slit gave a 2-pixel
resolution of 0.7\AA\ in the range $\lambda\lambda$6540-7170\AA. INT spectra were bracketed with Cu-Ne lamp
exposures and extracted and wavelength calibrated using
standard tasks from the {\sc iraf} package.

Radial velocity (RV) template stars were observed during twilight on
each night.  Spectrophotometric and telluric standards
were observed to allow for relative flux-calibrations and
telluric correction of the INT spectra. 
Heliocentric 
RVs were calculated by cross--correlation with template stars using the
{\sc fxcor} procedure in {\sc iraf}.  RV uncertainties were obtained by
combining standard errors in the RV measurements (from
multiple wavelength regions and multiple exposures) with systematic
uncertainties estimated from multiple measurements of RV standards.


Li\,{\sc i}~6708~\AA\ and H$\alpha$ equivalent widths (EWs) were 
measured relative to a
pseudo-continuum using the \,{\sc splot} procedure in \,{\sc iraf}.  No
attempt was made to deblend Li from the weak (EW$\sim 20$\,m\AA) 
neighbouring Fe line
at 6707.4~\AA. EW uncertainties were
approximated as $1.6\times\sqrt{\textrm{FWHM}\times{p}}/\textrm{SNR}$,
where FWHM, $p$ and SNR are the full-width half maximum of the measured line
and pixel size (in \AA), and the signal-to-noise ratio respectively
\citep{1988Cayrel}.  If no Li feature was seen, an upper limit was
estimated as twice this uncertainty assuming a FWHM of 0.7~\AA.

Spectral-types were determined from the TiO molecular band flux ratio 
$f(\lambda\lambda$7125-7136\AA)\,/$f(\lambda\lambda$7042-7046\AA).
This ratio is calibrated for spectral types of
K5--M7 \citep{1997Gizis}.
A comparison of our spectral types versus $V-K$ colour
with those of known BPMG
members with spectral types 
from \cite{2004Zuckerman_Song} and \cite{2013Malo_et_al},
suggests our calibration is consistent with literature values and that
the precision is about half a subclass.

\section{Analysis}
\begin{figure}
\subfigure{\includegraphics[width=0.48\textwidth]{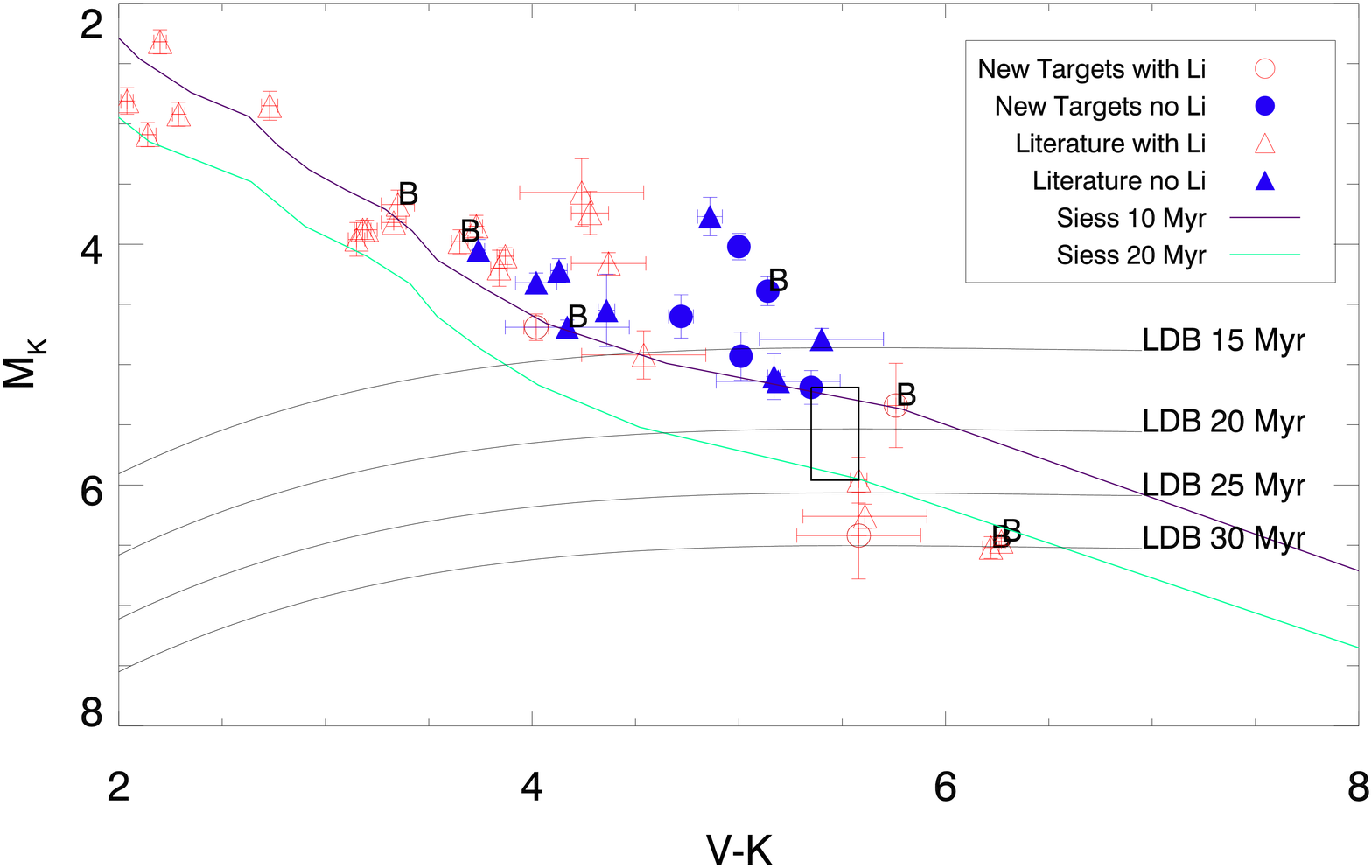}}
\subfigure{\includegraphics[width=0.48\textwidth]{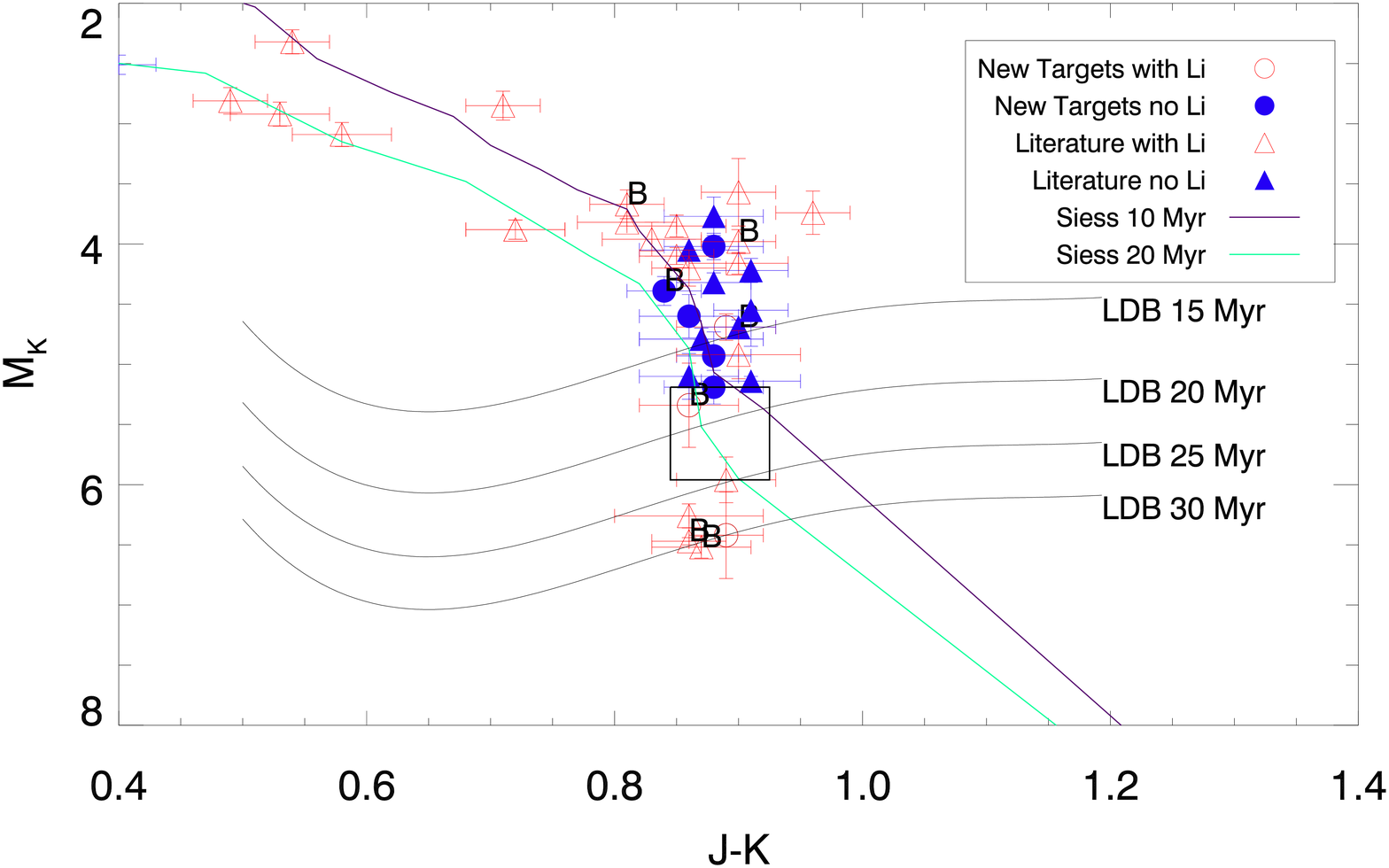}}
\subfigure{\includegraphics[width=0.48\textwidth]{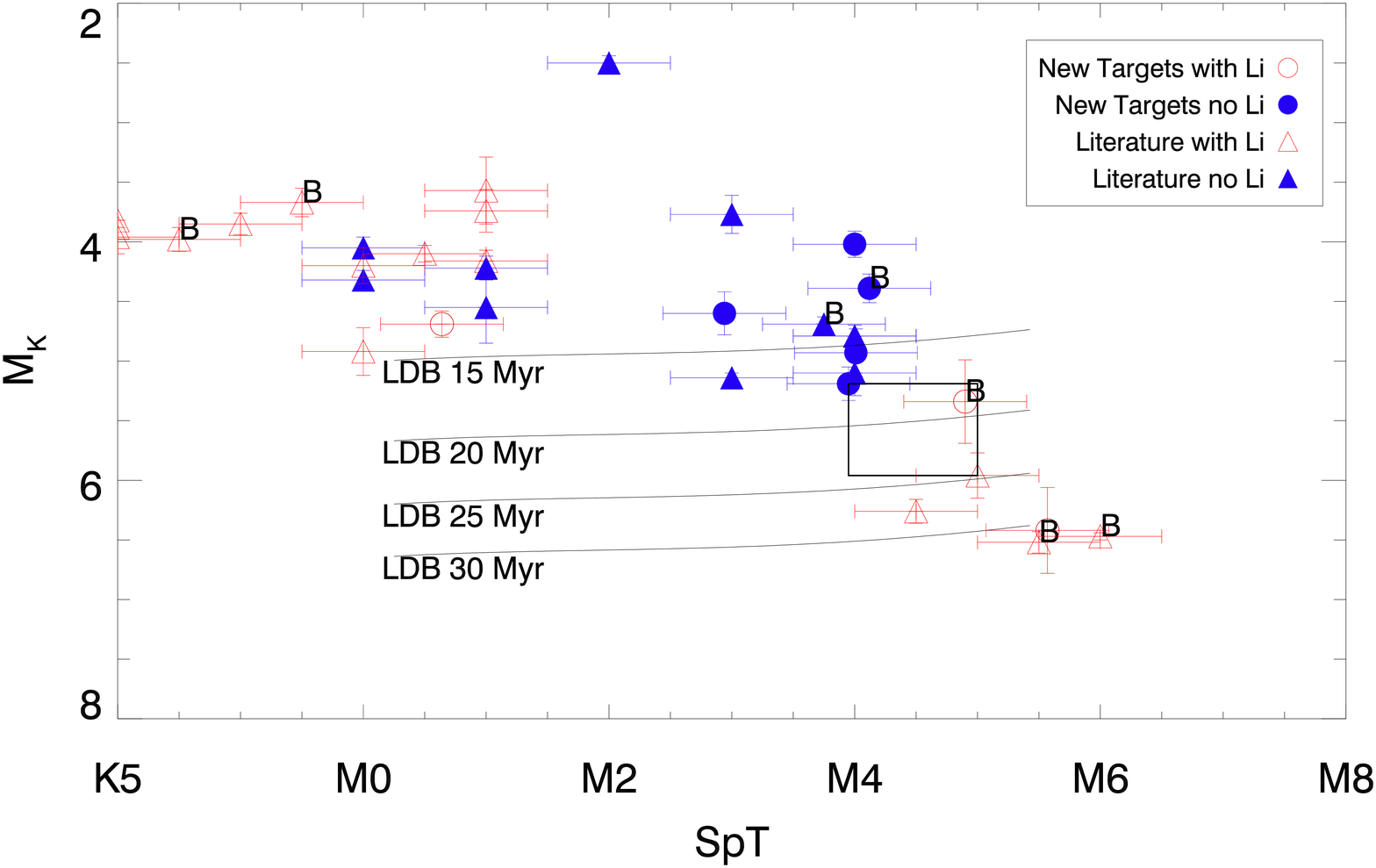}}
\centering
 \caption{Locating the LDB in 3 separate colour (or spectral-type)
   vs. magnitude diagrams. New members from Table~1 and objects from
   the literature are indicated. Absolute magnitudes are calculated from
   2MASS $K$ and a
   trigonometric parallax where available or a kinematic distance
   otherwise (see text). Known, unresolved binaries are marked with
   `B'. Black lines represent constant luminosity loci from
   \protect\cite{1997Chabrier_Baraffe} where Li is predicted to be $99\%$
   depleted at the ages indicated.  The green and maroon lines are 10
   and 20 Myr isochrones from \protect\cite{2000Siess_et_al}.
   The rectangle in each diagram represents the estimated LDB
   location and its uncertainty, based on the faintest Li-depleted member and the
  brightest Li-rich member (but excluding the unresolved binary at $M_K
  \simeq 5.3$, see Section 3.2).}

  \label{fig:CMDs}
\end{figure}

\subsection{Membership selection}

The mean Galactic space motion of the BPMG (U, V, W) = ($-$11, $-$16,
$-$9$\,\textrm{kms}^{-1}$) and 
convergent point (05h 19m 48s, $-$60d 13m 12s, Zuckerman \& Song 2004), 
leads to a predicted RV for group members
as a function of their sky position. For 8 candidates we found
the difference between this predicted RV and the measured RV
($\Delta$RV in
Table~1) was $<5$\,km\,s$^{-1}$ which, in common with previous
work \citep[e.g.,][]{2012Schlieder_et_al}, we adopt as a required membership criterion.
In addition, we require that an M-type member of the BPMG has
H$\alpha$ in emission, due to the strong chromospheric activity of such
young stars. Finally, the predicted tangential velocity of a BPMG
candidate combined with its proper motion, taken from the source paper
listed in Table~1, leads to a
``kinematic distance''. For 2 of the 8 candidates that pass the RV and
H$\alpha$ tests, there is a trigonometric parallax that agrees
reasonably well with this kinematic distance. For the rest, we demand
that when plotted on an absolute magnitude versus colour diagram using
the kinematic distance, that candidates lie on or above the
sequence defined by previously known members. Some latitude is allowed
above the sequence, because targets may be unresolved binaries
(two are known to be -- see Table 1)
up to 0.75 mag brighter than single stars of similar colour.


The colour-absolute--magnitude and spectral-type--magnitude diagrams are
shown in Fig.~\ref{fig:CMDs}. Our probable BPMG members are supplemented with
previously known members
\citep[from][]{2006Torres_et_al, 2008Mentuch_et_al, 2010Yee_Jensen, 2013Malo_et_al}, 
all of which also satisfy
the RV criterion discussed above. All 8 candidates that pass the
RV and H$\alpha$ tests are also consistent with the BPMG
sequence and we will assume they are BPMG members.


\subsection{Locating the LDB}

In M-dwarfs, the Li~{\sc i}~6708\AA\ feature is expected to be strong
where no depletion has occured (EW $\simeq 0.6$\AA,
Palla et al. 2007) and is observed to have this strength in very
young T-Tauri stars \citep[e.g.,][]{2013Sergison_et_al}. Li-burning should begin at spectral
types M2-M3 after $\sim 10$\,Myr. Within a few Myr these stars should
deplete Li by factors
$>100$, resulting in an EW $<0.2$~\AA, and Li burning progresses
towards cooler, less luminous stars. The most
model-independent way to define the LDB age (see Jeffries 2006), is to find the {\it luminosity} where M-type 
BPMG members
make the transition from having depleted their Li by factors $>100$,
to having undepleted Li at only slightly lower luminosities.

\nocite{2007Palla_et_al}
\nocite{2006Jeffries}

This is illustrated in Fig.~\ref{fig:CMDs}, where we show the absolute
$K$ magnitudes of Li-rich and
Li-poor (EW $<0.2$~\AA ) objects as a function of colour or spectral
type.  Overplotted are curves of constant luminosity corresponding to
the LDB (99 per cent Li depletion) at several ages. These are
obtained from theoretical models \citep{1997Chabrier_Baraffe} and
transformed to absolute $K$ magnitudes using quartic relationships fitted to tables of bolometric
corrections for PMS stars as a function of colour or spectral type \citep{2013Pecaut_Mamajek}. In
each diagram there is a reasonably clear boundary between Li-rich and
Li-poor BPMG members. We identify the LDB as a rectangular region
separating Li-poor from Li-rich stars. The upper bound is defined by
the faintest Li-depleted star, the lower bound is defined by the
brightest Li-rich star (excluding J05241914-1601153,
which is known to be an unresolved binary and will be brighter than a
single star at the same abcissa value -- see Table~1). The width of
the box is defined by the separation of these two stars, or twice their
average uncertainty, whichever is larger.

\begin{table*}
\centering
\caption{LDB ages for the BPMG. Each column corresponds to a
  diagram in Fig.~2; each row gives ages based on a different evolutionary
model.}

\begin{tabular}{lccc}
\toprule
	& $M_{K}$ vs V$-$K		& $M_{K}$ vs J$-$K		& $M_{K}$ vs SpT \\
\midrule

LDB location				& $M_{K}$ = 5.575 $\pm$ 0.385	& $M_{K}$ = 5.575 $\pm$ 0.385	& $M_{K}$ = 5.575 $\pm$ 0.385 \\
					& V$-$K = 5.465 $\pm$ 0.115	& J$-$K = 0.885 $\pm$ 0.040	& SpT = 4.475 $\pm$ 0.525 \\
$M_{\textrm{bol}}$ & 8.280 $\pm$ 0.544 & 8.321 $\pm$ 0.556 & 8.307 $\pm$ 0.546 \\
\midrule
Ages (Myr) &  &  & \\
\cite{1997Chabrier_Baraffe} & $20.3_{-3.2}^{+3.7}$	& $20.7_{-4.1}^{+4.6}$	& $20.6_{-3.2}^{+3.8}$ \\
\cite{2000Siess_et_al}  & $19.9_{-3.7}^{+4.1}$	& $20.5_{-4.9}^{+5.0}$	& $20.2_{-3.8}^{+4.1}$ \\
\cite{2004Burke_et_al} & $18.5_{-3.0}^{+3.8}$	& $18.9_{-3.8}^{+4.6}$	& $18.8_{-3.0}^{+3.8}$ \\
\bottomrule
\end{tabular}
\label{table:BPMG_Members3}
\end{table*}




The LDB age is calculated from the position of the central points of the
boxes in Fig.~2 by interpolating the LDB curves.
Age uncertainties are estimated by perturbing the abcissae
and ordinates according to the height and width of the LDB boxes and
adding the resultant age perturbations in quadrature. We also include
(in quadrature) an additional $\pm 0.1$ mag of uncertainty in colour
and magnitude and $\pm 0.5$
subclasses in spectral type to reflect likely errors in the calibration of
these quantitities.


The results are given in the first row of Table~2 for the models of
\cite{1997Chabrier_Baraffe}. To gain insight into any model
dependency we repeated the process using the models of
Siess et al. (2000, models with metallicity $Z=0.02$) and
\cite{2004Burke_et_al}. Table~\ref{table:BPMG_Members3} shows that
there is only $\sim$ 2 Myr between the youngest and oldest age
estimates from these models (compare the ages in a column). A
comparison of the ages in each row shows differences of $\leq 0.6$\,Myr,
attributable to small (0.1 mag) differences in the LDB location in
each diagram; the applied bolometric corrections are similar to $\leq
0.04$~mag. Finally, we note that defining the LDB as the luminosity at
which Li is depleted by 99.9 or 90 per cent would only change the
age estimates by $\pm 1$\,Myr. All these uncertainties are small compared
with the 3--4\,Myr observational error due to the size of the estimated
LDB boxes.  Adopting the Chabrier \& Baraffe models, our final LDB age
estimate is $21 \pm 4$\,Myr, with an additional
model dependent uncertainty of only $\pm$ 1 Myr.

\section{Discussion and implications}

The LDB age derived for the BPMG is reasonably precise, but
could be improved by the addition of data for more members close to the
LDB and an accurate assessment of their binary nature. More importantly,
the LDB method yields an age likely to have a high degree of
absolute accuracy. The physics involved in calculating the luminosity
at the LDB versus age is well understood. Numerical experiments
adjusting the physical inputs of models (convection efficiency, opacities,
equation of state etc.) within plausible bounds yield age uncertainties
of only 8 per cent at $\sim 20 $\,Myr \citep{2004Burke_et_al},
comparable with the model dependence we have found.

Previous work on the LDB in the BPMG concluded that
the age was $>30$\,Myr, incompatible with ages derived from
fitting isochrones in the HRD or from kinematic traceback, and suggesting
there could be gaps in our understanding of the physics of Li
depletion \citep{2002Song_et_al, 2010Yee_Jensen}. Both of these works
suffered from a sparse sample (only the binary pair HIP112312AB was
considered by Song et al. 2002); the new BPMG members confirmed here
locate the LDB with more precision, particularly in defining the
lowest luminosity objects that have lost their Li. However, all the
objects considered in previous work are present in Fig.~2 and are
entirely consistent with our LDB age. Even if we ignore the
new members claimed in this {\it Letter}, the presence of
previously known Li-depleted objects at $M_K > 5.0$ constrains the 
LDB age to be $>15$\,Myr. We think the main reason that 
previous work found an older age was 
that comparison was made with Li depletion models as a
function of $T_{\rm eff}$. This is a far more uncertain enterprise than
comparing the {\it luminosity} of the LDB with
models. Measuring $T_{\rm eff}$ in low-mass stars 
has systematic uncertainties of 100-200\,K at
best, leading to large LDB age errors because the Hayashi
tracks of stars with different mass are close together in $T_{\rm eff}$. 
Furthermore, theoretical $T_{\rm
  eff}$ predictions are extremely sensitive to adopted convection
efficiencies and atmospheres, making any age estimate highly model-dependent.

The LDB age we find is at the upper end of age estimates from 
isochronal fits to low mass BPMG members in the HRD --
$12^{+8}_{-4}$\,Myr (Zuckerman et al. 2001). Isochronal ages are 
also model dependent, are very sensitive to adopted convective
efficiencies, can vary depending on 
which mass range is considered, may
be biased downwards by the neglect of unresolved binarity and
also depend on how colours and spectral types
are translated into $T_{\rm eff}$ for comparison with models (or vice
versa). The situation is perfectly illustrated in the top panels of
Fig.~2, where a single isochrone
is incapable of fitting all the low-mass members. Nevertheless,
\citet{2013Jeffries_et_al} and \citet{2013Bell_et_al} have recently shown that
for NGC~1960, a rich open cluster with a similar LDB age to the BPMG,
that isochronal fits to both low- and high-mass stars do agree with
the LDB age when these problems are carefully considered; the same
may yet be true for the BPMG.

In principle, kinematic traceback ages provide a model-independent age,
or at least the time since the MG occupied a minimum spatial
extent. Our LDB age is older than traceback ages of
10--12\,Myr reported by \cite{2002Ortega_et_al,2004Ortega_et_al} and \cite{2003Song_et_al}
, which have formal uncertainties as small as
0.3\,Myr. However, other kinematic analyses do not concur with this age
or this precision. \cite{2006Torres_et_al} find a group expansion consistent with an age of $\sim
20$\,Myr; \cite{2007Makarov_et_al} give a time of closest approach for pairs of
BPMG members as $22\pm 12$\,Myr ago; and 
Mamajek (2013, private communication), using the
revised Hipparcos astrometry, was unable to find any significant evidence that
BPMG was smaller in the past. The differing conclusions appear to arise
from uncertain space motions combined with some
subjectivity in which group members are included in the
analyses. The forthcoming Gaia astrometry mission may reveal a
precise kinematic age for the BPMG, but for now it appears an
unreliable technique.

A key role for BPMG members is in testing evolutionary models for
low-mass objects and circumstellar material at young ages.
Adopting an older age of 21\,Myr for the BPMG changes the inferred
masses of substellar and planetary companions. Biller et al. (2010)
find a faint companion to the BPMG member PZ~Tel, estimating a mass for
PZ~Tel~b of 36\,$M_{\rm Jup}$ at an assumed age of 12\,Myr. An age of
20\,Myr would increase the inferred mass by $\sim 30$ per cent. 
Similarly, based on an age of $12^{+8}_{-4}$\,Myr, Bonnefoy et al. (2013) estimate a mass of $10^{+3}_{-2}\,M_{\rm Jup}$ for
$\beta$~Pic~b, the
uncertainties largely arising from the assumed age. Again, an increase
in age to 21\,Myr results in a $\sim 30$ per cent increase in
inferred mass, which is however still below the upper limit of
$15.5\,M_{\rm Jup}$ currently imposed by dynamical constraints
\citep{2012Lagrange_et_al}.

\section*{Acknowledgements}
Based on observations made with the Nordic Optical Telescope (46-102),
operated by the Nordic Optical Telescope Scientific Association and
with the Isaac Newton Telescope (I/2013A/1) operated by the Isaac Newton
Group, at the Spanish Observatorio del Roque de los Muchachos of
the Instituto de Astrofisica de Canarias.  The research
leading to these results has received funding from the European Union
Seventh Framework Programme (FP7/2007-2013) under grant agreement
No. 312430 (OPTICON).  This research has made use of the SIMBAD
database, operated at CDS, Strasbourg, France. ASB acknowledges the
support of the STFC.
\bibliography{BPMG}

\label{lastpage}
\end{document}